# *Mini array of quantum Hall devices based on epitaxial graphene*

S. Novikov,[1] N. Lebedeva,[1] J. Hämäläinen,[2] I. Iisakka,[2] P. Immonen,[2] A.J. Manninen,[2] and A. Satrapinski[2]

[1]*Department of Micro and Nanosciences, Aalto University, Micronova, Tietotie 3, Espoo, Finland*

[2]*VTT Technical Research Centre of Finland Ltd, Centre for Metrology MIKES, P.O. Box 1000, 02044 VTT, Finland*

Series connection of four quantum Hall effect (QHE) devices based on epitaxial graphene films was studied for realization of a quantum resistance standard with an up-scaled value. The tested devices showed quantum Hall plateaux $R_{H,2}$ at filling factor $v = 2$ starting from relatively low magnetic field (between 4 T and 5 T) when temperature was 1.5 K. Precision measurements of quantized Hall resistance of four QHE devices connected by triple series connections and external bonding wires were done at $B = 7$ T and $T = 1.5$ K using a commercial precision resistance bridge with 50 µA current through the QHE device. The results showed that the deviation of the quantized Hall resistance of the series connection of four graphene-based QHE devices from the expected value of $4 \times R_{H,2} = 2h/e^2$ was smaller than the relative standard uncertainty of the measurement ($< 1 \times 10^{-7}$) limited by the used resistance bridge.

**I. INTRODUCTION**

Graphene has been in focus of electrical metrology ever since its discovery [1] and especially after the quantum Hall effect (QHE) was observed in graphene in 2005 [2,3]. In QHE, the Hall resistance of a two-dimensional (2D) system of charge carriers as a function of magnetic field or charge carrier density shows well-defined quantized plateaus at $R_{H,v} = R_K/v$, where $R_K = h/e^2$ is the von Klitzing constant, $h$ is the Planck constant, $e$ is the elementary charge, and $v$ is an integer [4]. Quantized Hall resistance (QHR) has been used as a primary standard for resistance in national metrology institutes since the beginning of 1990's with a reproducibility that is about two orders of magnitude better than the uncertainty of the determination of the ohm in the present international system of units SI [5]. The importance of QHE for metrology will rise to a new level in a major revision of the SI, possibly in 2018, after which the base units of SI will be defined by fixing the numerical values of some fundamental constants [6]. Then the QHR standards can be used for direct realization of the "new ohm" via the fixed values of $h$ and $e$.

The reason for metrologists' great interest on graphene-based QHE devices is that they can be operated in much more easily achievable experimental conditions than the conventional QHR standards based on GaAs/AlGaAs heterostructures, which typically require a very high magnetic field $B > 10$ T and a low temperature $T < 1.5$ K for accurate measurements [7]. This is a consequence of the peculiar nature of charge carriers in graphene. They behave as massless Dirac fermions and have a linear dispersion relation, instead of the parabolic dispersion relation and finite effective mass of charge carriers in semiconductor-based 2D systems such as GaAs/AlGaAs heterostructure. Due to this, the spacing between energy levels,

Landau levels, is much larger in graphene than in conventional 2D systems, and QHE can be observed in relaxed experimental conditions. In graphene, QHE has been observed even at room temperature [8], and accurate QHR has been measured at temperatures up to 10 K and in magnetic fields down to 2.5 T [9,10,11].

In both GaAs-based and graphene-based QHR standards, the quantized level $R_{H,2} = R_K/2 \approx 12.906$ k$\Omega$ is used as the reference value in accurate resistance measurements. Other reference values can, at least in principle, be obtained by series or parallel connection of individual Hall devices [12]. Multiple-connected arrays of GaAs-based Hall devices have already been developed and used successfully for accurate scaling of dc resistances in the range 100 $\Omega$ – 1.29 M$\Omega$ [13,14,15,16], but they are not in widespread use due to difficulties of reproducible fabrication of accurate and reliable devices with hundreds of contacts between individual Hall bars that should have very identical characteristics. In addition to the relaxed experimental requirements of graphene QHE devices compared to the GaAs ones, it is expected that many problems of GaAs-based quantum Hall array resistance standards (QHARS) could be solved more easily in graphene-based devices [7]. Fabrication of the first prototype of a graphene-based QHARS with 100 Hall bars connected in parallel was reported recently [17], and it was shown that the quantized resistance of the device matched the expected value of $R_{H,2}/100 \approx 129.06$ $\Omega$ within the relative measurement accuracy of about $10^{-4}$ in magnetic fields between 7 T and 9 T with a low measurement current of 1 µA through each Hall bar. Also, the first experimental feasibility study of an interconnect-less QHARS that utilizes the unique possibility of graphene to have both n-and p-type charge carriers in the same device has been reported [18]: a Hall bar with a p-n junction in the middle was fabricated using exfoliated graphene, and the four-terminal resistance between voltage terminals at n- and p-type regions showed a clear plateau as a function of gate voltage at $R \approx 2 \times R_{H,2} \approx 25.8$ k$\Omega$, as expected. However, no experiments on graphene-based QHARS devices at metrologically relevant uncertainty level below $10^{-7}$ have been reported before our work.

There are many important needs for accurate and reliable QHARS devices in metrology. Parallel-connected low-value QHARS devices can be operated with high current to decrease uncertainty, and the current level would be suitable to allow their use as a reference in a conventional resistance bridge, too. Parallel-connected QHARS references would enable direct calibration of e.g. the metrologically important 1 $\Omega$ resistance standard against the quantized Hall resistance. Series-connected QHARS references could be used for direct calibration of high value resistance standards without use of resistive scaling devices. Series-connected QHARS devices would also be needed in ac calibration of impedance standards. In recent experiments with single graphene-based Hall devices, the QHE plateaux measured with alternating current were found to be flat within one part in $10^7$, which is much better than for plain GaAs quantum Hall devices [19]. By using a QHARS instead of a single QHR device, the quantized resistance value can be tailored to a level that is suitable for direct use as a reference in quadrature bridge for calibration of capacitance standards. Most suitable resistance values for a QHARS that would be used for realization of capacitance unit from QHE are in the range 50 kΩ - 100 kΩ that can be realized by connecting e.g. 4 or 8 individual QHE devices in series [20]. The resistance of such an array is close to the impedances of most accurate capacitance standards at conventional frequency range 1 kHz – 10 kHz. Using such a QHARS array as the ac reference resistor in a quadrature bridge can replace the long chain of resistance/capacitance measurements by direct comparison of impedances of the capacitance standard and QHARS.

In this paper, we report on fabrication and precision measurements of a mini array of four quantum Hall devices based on epitaxial graphene. The Hall devices which were all fabricated on the same chip were connected using triple series connection [12] and external bonding wires. Results of magneto-resistance measurements of the four Hall devices showed quantized plateaus of the Hall resistance starting at a relatively low magnetic field and allowed to perform the first precision measurement on a graphene-based QHR array.

## II. FABRICATION OF GRAPHENE DEVICES

Epitaxial graphene film was grown on Si face of 4H-SiC substrate by annealing in Ar ambient at atmospheric pressure and temperature near 1700 °C for 5 minutes. Details of the film fabrication technology are given in Refs. [9,21]. The film thickness was estimated by means of Auger spectroscopy that confirmed the presence of a single layer of graphene before patterning. However, we cannot exclude the possibility of formation of islands with double-layer graphene in a large-area epitaxial film [22].

Patterns for the Hall bars and the contacts were made using laser photolithography with AZ5214 resist. Reactive ion etching in argon–oxygen plasma was applied to remove the graphene layer from uncoated areas. Four QHE devices with the channel size 1000 μm × 200 μm were fabricated on one 5 mm × 5 mm chip. An example of one of the chips with four Hall devices is shown in Fig. 1 (a). For fabrication of reliable and low-resistance contacts, a two-step metallization process [23] was used. Double metal–graphene contacts were made by e-beam evaporation and lift-off photolithography. In the first step, Ti/Au (5/50 nm) was used, and in the second, Ti/Au (200/300 nm) metallization was used for reducing the contact resistance. The first metallization was made on the SiC surface areas from where the graphene film had been etched away.

The carrier density $n_c$ was controlled by measuring the square resistance on corresponding terminal pairs when 10 μA current passed through all four devices. The chip with Hall devices was covered with two polymers, PMMA and ZEP500, which allow adjusting $n_c$ by photochemical gating [24]. In our technological process, exposition of the prepared sample in hot air before applying photochemical gating is used [25]. In order to introduce additional oxygen related doping centers into the sample that we used in our metrological experiments, exposition in hot air at 120 °C was performed for about 1 hour just after the fabrication of the Hall bars and the contacts but before covering with bilayer polymer [9]. This gave an estimated initial carrier density in the range $(1.6 - 3.1) \times 10^{11}$ cm$^{-2}$, and the mobility was estimated to be about $(1400 - 2700)$ cm$^2$V/s. Illumination with UV light was not used to control the carrier density of the devices of this report.

The scheme of the interconnections for a series array of four Hall devices, with the voltage terminals connected in multiple series connection [12] using external bonding wires is presented in Fig. 1 (b). Each of the four Hall bars on the chip has the possibility for independent connection and measurement of the Hall voltage at potential terminals on the sides of the Hall bar. Additional cross connections serve to suppress the influence of the non-negligible contact resistance of interconnections on the serial Hall voltage [14,15]. These connections have been made using Al bonding wire with the resistance of each wire about 10 mΩ.

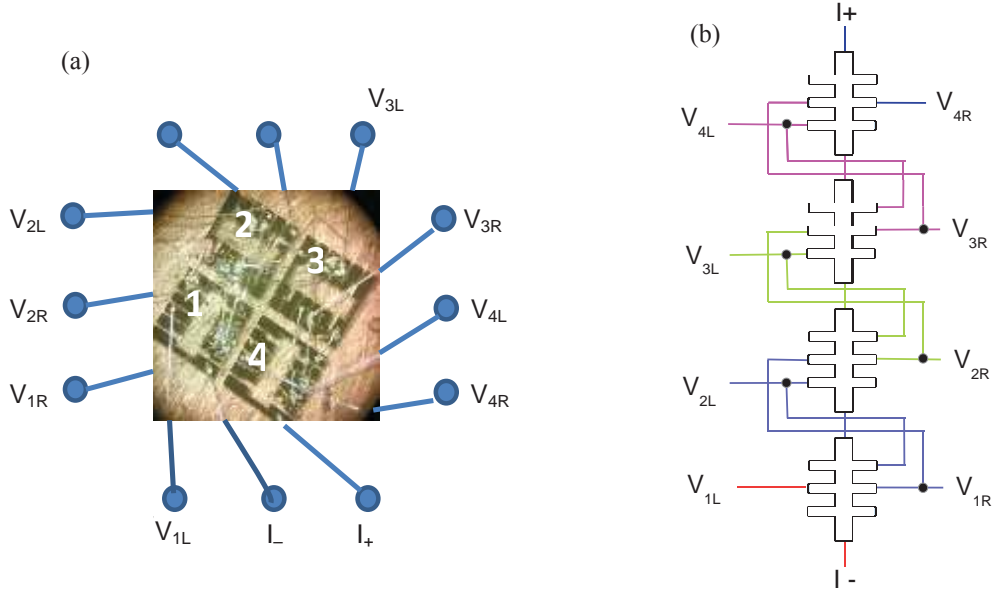

FIG. 1. (a) A photograph of the chip with four quantum Hall devices connected in series using multiple series connection and external bonding wires. (b) Scheme of configuration of four Hall bars with the current and voltage terminals connected for multiple series connection.

## III. EXPERIMENTAL RESULTS AND DISCUSSION

### A. Breakdown current

An 11 T superconducting magnet system with variable temperature insert and with temperature control down to 1.5 K was used for the measurements. The QHE chip was cooled down to 1.5 K slowly, during two hours. In the multiple series connection of Fig. 1 (b), the longitudinal resistance $R_{xx}$ of individual Hall bars cannot be measured directly. Instead, we applied a dc current $I_{sd}$ between terminals I₋ and I₊ (see Fig. 1 (b)) and measured the dc voltage $V_b$ between potential terminals on the same side of the Hall bar (e.g. between terminals $V_{1R}$ and $V_{2L}$) using a precision multimeter. Measured resistance $R_b = V_b/I_{sd}$ is proportional to the difference of the longitudinal resistances, $R_{xx}$ of two multiple-connected Hall bars (e.g. Hall bars 1 and 2). When the breakdown current of one of those Hall bars is exceeded, $R_b$ starts deviating from zero.

One of the criteria for the reliability and correctness of the QHR standard is that $R_{xx}$ must be unmeasurably small in magnetic fields corresponding to the quantized plateaus of the Hall resistance $R_{xy}$. We measured $R_b$ using $I_{sd} = 10$ µA, and its general dependence on magnetic field between 0 T and 7 T is presented in Fig. 2 for three terminal pairs (see Fig. 1): $V_{1R}$ - $V_{2L}$ (Hall bars 1 and 2), $V_{2R}$ - $V_{3R}$ (Hall bars 2 and 3), and $V_{3R}$ - $V_{4L}$ (Hall bars 3 and 4). All terminal pairs show a rather peculiar behavior at low magnetic fields, including changes in the sign of the effective resistance, and different values of $R_b$ at $B = 0$. The origin of this is not fully understood yet, but such behavior can be caused by inhomogeneities in the thickness of the graphene film, presence of bilayer domains, and by the peculiarities of the triple series connection of Hall bars with non-identical characteristics. However, in spite of these issues, the measured $R_b$ between all tested terminal pairs approaches zero when the quantized plateaus of Hall resistance are reached at $B > 6$ T, proposing that the longitudinal resistances of all Hall bars are close to zero in QHE conditions.

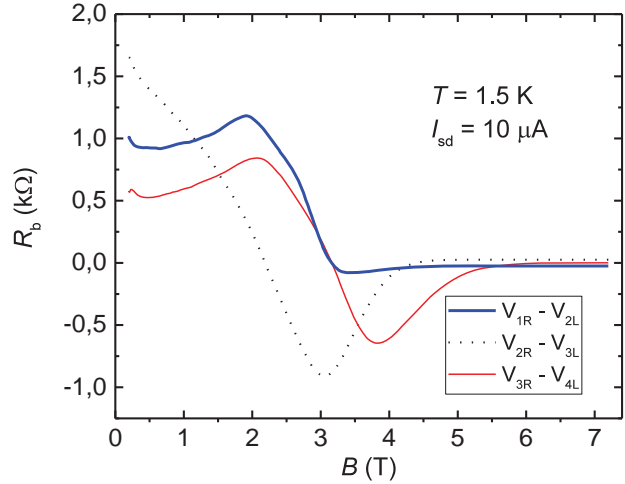

FIG. 2. Magnetic field dependence of resistance $R_b$ (see the text) of the following Hall bars according to the connection scheme of Fig. 1: Hall bars 1 and 2 (thick blue line), Hall bars 2 and 3 (dotted black line), and Hall bars 3 and 4 (thin red line). Terminal pairs used in measurements are indicated in the legend using the notation of Fig. 1. Measurements were done at $T = 1.5$ K with measurement current $I_{sd} = 10$ µA. The apparent deviation of $R_b$ from zero at $B > 6$ T on this resistance scale is caused by non-compensated thermal and offset voltages of the measurement system.

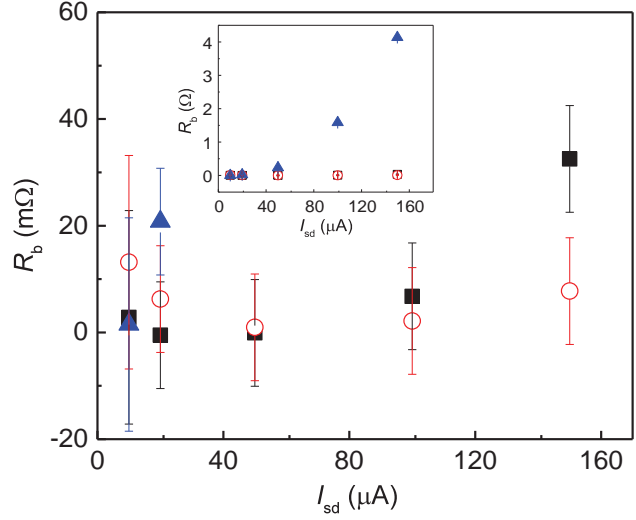

FIG. 3. Measurement of breakdown currents. Current dependence of resistance $R_b$ (see the text) of the following Hall bars according to the connection scheme of Fig. 1 is presented: Hall bars 1 and 2 using terminals $V_{1R} - V_{2L}$ (open red spheres), Hall bars 2 and 3 using terminals $V_{2R} - V_{3L}$ (solid black squares), and Hall bars 3 and 4 using terminals $V_{3R} - V_{4L}$ (solid blue triangles). The inset shows current dependence of $R_b$ on a larger resistance scale. Measurements were done at $T = 1.5$ K and $B = 7$ T.

Figure 3 shows results of more accurate measurements of $R_b$ as a function of measurement current $I_{sd}$ at magnetic field $B = 7$ T, which corresponds to the quantized Hall resistance plateau $R_{xy} = R_{H,2} = R_K/2$ for these devices. The same terminal pairs were used as in the experiments of Fig. 2. In order to eliminate thermal and offset voltages, the polarity of $I_{sd}$ was alternated at intervals of about 20 s. As Fig. 3 shows, the measured values of $R_b$ for Hall bars 1 and 2 (open red spheres) and Hall bars 2 and 3 (solid black squares) do not exceed the resolution of $R_b$ measurement (about 10 m$\Omega$) when $I_{sd} < 100$ µA. However, a significant increase in $R_b$ of Hall bars 3 and 4 (solid blue triangles) is observed for $I_{sd} > 50$ µA (see the inset in Fig. 3), which indicates that the breakdown current in Hall bar 3 or 4 is exceeded. Nevertheless, the results indicate that it should be possible to achieve accurate quantization of QHE in all of the four Hall bars if a low enough measurement current not exceeding 50 µA is used.

## B. QHR measurements of four Hall devices connected in series

Figure 4 shows the magnetic field dependence of the Hall resistance, $R_{xy}$, corresponding to a series connection of $N = 1$, 2, 3, or 4 Hall bars. Using the scheme of Fig. 1, a dc current $I_{sd} = 10$ µA was applied between terminals $I_-$ and $I_+$, and the Hall voltages were measured between terminals $V_{1L} - V_{1R}$ ($1 \times R_{H,2}$), $V_{1L} - V_{2R}$ ($2 \times R_{H,2}$), $V_{1L} - V_{3R}$ ($3 \times R_{H,2}$), and $V_{1L} - V_{4R}$ ($4 \times R_{H,2}$). It is seen that in all four Hall terminal pairs, quantization to level $R_{xy} = N \times R_{H,2}$ is observed at magnetic fields above about 4 or 5 T. In lower magnetic fields, the Hall resistance shows a nonideal behavior and $R_{xy}$ has a large value of more than 10 k$\Omega$ per Hall bar even at $B = 0$ (i.e. there is a perpendicular voltage across the Hall bars even at $B = 0$). We have earlier observed a similar behavior in some experiments with single graphene-based Hall bars, too. The exact reason of this is not presently known, but the peculiarities in the low-field behavior of $R_{xy}$ and $R_b$ can at least partly be caused by the same reasons.

Epitaxial single layer graphene always contains small fraction of double layer graphene [26, 27]. Double layer graphene exists in the form of islands elongated along atomic steps. The conductivity of double layer graphene becomes significant for the total conductivity when the carrier concentration is very low. Since orientation of the Hall channel along atomic steps is not ideal, double layer islands can be oriented in diagonal to the current channel and finally generate potential on the Hall contacts without magnetic field. At very low carrier density, it may cause local changes in the spatial distribution of the current flow and/or influence of geometrical inaccuracy of the potential contacts of left and right Hall bars edges, and that could cause appearance of the voltages across the Hall bars even when no external magnetic field is applied.

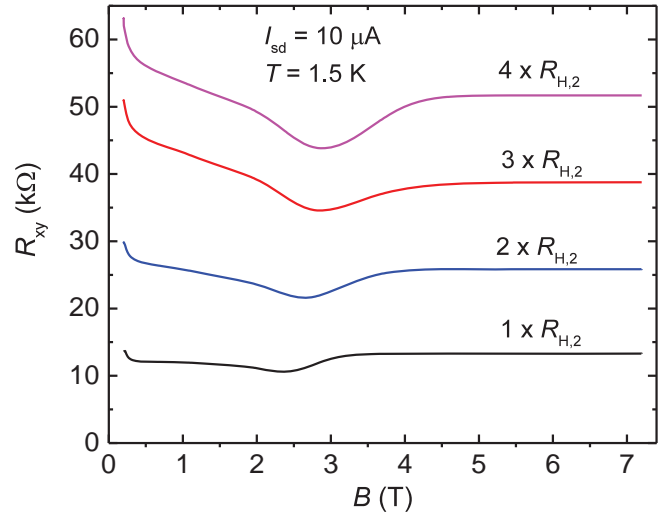

FIG. 4. Magnetic field dependence of the transverse resistance measured at $T$ = 1.5 K between terminals $V_{1L}$ - $V_{1R}$ (1×$R_{H,2}$), $V_{1L}$ - $V_{2R}$ (2×$R_{H,2}$), $V_{1L}$ - $V_{3R}$ (3×$R_{H,2}$), and $V_{1L}$ - $V_{4R}$ (4×$R_{H,2}$) with a dc current $I_{sd}$ = 10 µA applied between terminals I- and I+ (see Fig. 1).

As the main result of this paper, Fig. 5 shows an indirect comparison of the mini array of four series-connected graphene QHR devices and a GaAs-based QHR standard. A stable standard resistor with nominal value $R_0$ = 10 kΩ and real value $R_{DUT}$ = $R_0$ + Δ$R$ was calibrated against two references: first against the array of four graphene QHR devices and then against a conventional resistance standard that has traceability to a GaAs-based QHR standard. Results in Fig. 5 are expressed as the relative deviation Δ$R$/$R_0$ of $R_{DUT}$ from its nominal value $R_0$ = 10 kΩ. Consistency between calibration results against the two different references indicates consistency between the references.

Measurements were done using a resistance bridge MI 6242B, whose specified relative uncertainty for the ratio of two resistances in the range 100 Ω – 10 kΩ is better than 1×10$^{-7}$ when nominal current is used. Black squares in Fig. 5 show results of 5 calibration runs (total duration about 18 hours) of the 10 kΩ resistor, when the QHR of 4 series-connected Hall bars of Fig. 1 was used as the reference and bridge ratio was 5.16:1. The relative deviation of the 10 kΩ resistor from its nominal value was determined from the bridge measurement results assuming an exact value $R = 4 \times R_{H,2} = 2h/e^2$ (≈ 51.625 kΩ) for the Hall resistance of the graphene-based QHR array. The dc current through the QHR device was 50 µA, which corresponds to 258 µA current through the 10 kΩ resistor under calibration. This is somewhat smaller than the nominal current of the bridge (1 mA in measurement of a 10 kΩ resistor against a 10 kΩ reference), but higher currents could not be used due to the limitations of the QHR device (see Fig. 3).

Red spheres in Fig. 5 are calibration results of the same 10 kΩ resistor when a conventional 1 kΩ resistance standard was used as the reference and bridge ratio was 1:10. The 1 kΩ reference is traceable to a GaAs-based QHR standard via two calibration steps. First, a 100 Ω resistance standard was evaluated with respect to $R_{H,2}$ of a GaAs/AlGaAs QHR standard using a Cryogenic Current Comparator (CCC) resistance bridge [28] with combined uncertainty of about $1 \times 10^{-8}$ ($\sigma = 1$). Then the 1 kΩ resistor was calibrated against the 100 Ω standard using a DCC MI 6242B bridge, so that finally its value in terms of $R_{H,2} = (1/2)h/e^2$ of a GaAs/AlGaAs QHR standard was known with combined relative uncertainty of $6 \times 10^{-8}$ ($\sigma = 1$). The difference between the results of the two groups of calibrations of Fig. 5 is $(-3.5 \pm 8) \times 10^{-8}$. Consistency of the results within measurement uncertainty of $8 \times 10^{-8}$ indicates that the quantized Hall resistance of our mini array of four series-connected QHE devices based on epitaxial graphene has the expected value $R_{xy} = 4 \times R_{H,2} = 2h/e^2$ with relative uncertainty of $8 \times 10^{-8}$ ($\sigma = 1$). The latter equality is based on well-established assumption that $R_{H,2} = (1/2)h/e^2$ for the GaAs/AlGaAs QHR standard within our measurement uncertainty [29].

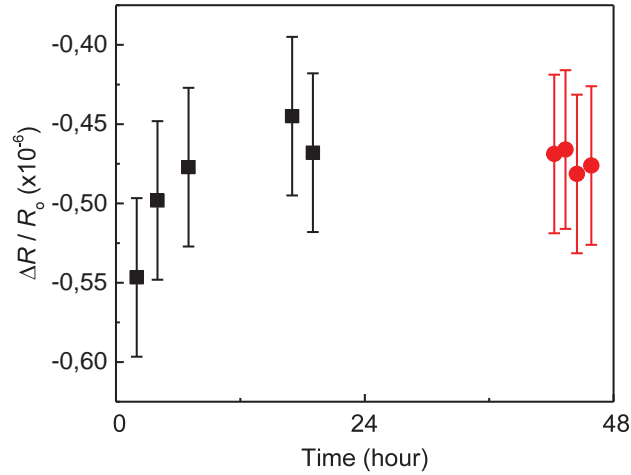

FIG. 5. Calibration results of a standard resistor with nominal value $R_0 = 10$ kΩ expressed as relative deviation $\Delta R/R_0$ from the nominal value (parts per million). In the first group of calibrations (black squares), the mini array of four series-connected graphene QHR devices with $R = 4 \times R_{H,2} = 2h/e^2$ was used as the reference. In the second group (red spheres), the reference was a conventional resistance standard with traceability to a GaAs-based QHR standard. Consistency between the calibration results indicates consistency between the references.

## IV. CONCLUSIONS

A mini array containing four QHE devices based on epitaxial graphene was fabricated and its performance as an up-scaled quantum standard of resistance was studied. All four Hall bars were fabricated on one chip and connected in triple series connection using external bonding wires. Indirect measurements indicated that the longitudinal resistance $R_{xx}$ between all tested contact pairs at $B = 7$ T and $T = 1.5$ K was below the measurement resolution of about 10 mΩ at very low current levels, but the resistance between one of the tested contact pairs increased rapidly when current exceeded about 50 μA. That set an upper limit for the measurement current in precision QHR measurements. In spite of this, it was possible to reach relative uncertainty of about of $8\times10^{-8}$ ($\sigma = 1$) in calibration of a 10 kΩ resistance standard by a commercial direct current comparator (DCC) resistance bridge, when the QHR of the series-connected array of four graphene-based Hall bars was used as the reference. This result was compared to that obtained when the reference was a conventional resistance standard that has traceability to a GaAs-based QHR, and the difference between the results was $(-3.5 \pm 8) \times 10^{-8}$. Consistency of the results within measurement uncertainty means that the quantized Hall resistance of the mini array of four series-connected QHE devices based on epitaxial graphene had the expected value $R_{xy} = 4\times R_{H,2} = 2h/e^2$ ($\approx 51.625$ kΩ) within relative uncertainty of $8\times10^{-8}$ ($\sigma = 1$).

This result demonstrates the feasibility of accurate graphene-based quantum Hall array resistance standards (QHARS), but it also indicates that developing a large array with tens or hundreds of Hall bars, each with breakdown current of 50 μA or more, is a challenging objective. Large arrays with on-chip interconnections would be needed in resistance scaling applications. The accurate mini-array of 4 or 8 Hall bars connected in series could be applied in ac impedance metrology to improve, e.g., traceability of 1 nF capacitance standards at frequencies near 1 kHz. Our experiments concentrated on dc properties of graphene-based QHR arrays, but recent ac measurements of the quantum Hall resistance in single graphene-based Hall bars demonstrated promising ac performance and low capacitive losses in the kHz range [19]. A major benefit of graphene-based QHARS devices compared to GaAs-based devices is that graphene-based devices can be operated in much more easily achievable experimental conditions: higher temperatures and lower magnetic fields. Our accurate experiments were performed at $T = 1.5$ K and $B = 7$ T, but at least some Hall bars of the array approached the quantized state already in magnetic fields below 5 T, and earlier experiments with single graphene-based QHE devices have demonstrated accurate QHR down to magnetic fields of 2.5 T. Full quantization in all Hall bars of the array in a low magnetic field would require a homogeneous charge carrier density that is lower than in the reported device. That is a challenging but feasible objective for further research towards a user-friendly scalable quantum resistance/impedance standard.